\documentclass[%
reprint,
superscriptaddress, 
amsmath,amssymb,
aps,
prl,
floatfix,
]{revtex4-2}

\usepackage{natbib} 
\usepackage{upgreek}
\usepackage{graphicx}
\usepackage{dcolumn}
\usepackage{bm}
\usepackage{hyperref}

\begin{document}
	\preprint{APS/123-QED}
	\hypersetup{
		hidelinks
	}
	
	\title{Generation of optical vortices imitating water vortices}
	
	\author{Jun Yao}
	\affiliation{School of Physics, University of Electronic Science and Technology of China, Chengdu 611731, China}
	
	\author{Yihua Bai}
	\affiliation{School of Physics, University of Electronic Science and Technology of China, Chengdu 611731, China}
	
    \author{Yaqiang Qin}
    \affiliation{State Key Laboratory of Molecular Developmental Biology, Institute of Genetics and Developmental Biology, Chinese Academy of Sciences, Beijing 100101, China}
    
    \author{Mingsheng Gao}
    \affiliation{School of Physics, University of Electronic Science and Technology of China, Chengdu 611731, China}
    
    \author{Lei-Ming Zhou}
    \affiliation{Department of Optical Science and
    Engineering, Hefei University of Technology, Hefei 230009, China}

    \author{Yuqiang Jiang}
    \email{yqjiang@genetics.ac.cn}
    \affiliation{State Key Laboratory of Molecular Developmental Biology, Institute of Genetics and Developmental Biology, Chinese Academy of Sciences, Beijing 100101, China}

	\author{Yuanjie Yang}
	\email{dr.yang2003@uestc.edu.cn}
	\affiliation{School of Physics, University of Electronic Science and Technology of China, Chengdu 611731, China}

	\date{\today}
	
	\begin{abstract}
	    In optics, we can generate vortex beams using specific methods such as spiral phase plates or computer generated holograms. While, in nature, it is worth noting that water can produce vortices by a circularly symmetrical hole. So, if a light beam can generate vortex when it is diffracted by an aperture? Here, we show that the light field in the Fresnel region of the diffracted circularly polarized beam carries orbital angular momentum, which can transfer to the trapped particles and make orbital rotation.
	\end{abstract}

	\maketitle
    In nature, vortices are ubiquitous, from tornadoes to water vortices. In hydraulic engineering, gravitational water vortex power plant, which use water vortices to generate electricity, can be used to provide power to remote areas\cite{paper01}. The study of water vortices can also be applied to protect hydraulic engineering\cite{paper02,paper03}. In aviation, the generation of the starting vortex can increase the lift of the aircraft\cite{paper04,paper05}. Optical vortices have also been intensively studied in recent decades due to their unique properties such as quantized orbital angular momentum (OAM). It is well known that the angular momentum (AM) of photons can be categorized as spin angular momentum (SAM) and OAM, which are related to its polarization states and helical wavefront, respectively\cite{paper06,paper07,paper08}. A light beam with a helical phase of exp($il\theta$) can carry an OAM of $l\hbar$ per photon, where $l$ is the topological charge and $\theta$ is the azimuthal angle\cite{paper09}. Such vortex beams possess a phase singularity at the beam center, resulting in a hollow intensity profile\cite{paper10}. Moreover, the OAM of vortex beams can be transferred to the particles, leading to the orbital rotation of particles around the beam axis\cite{paper11}. These characteristics make optical vortices widely used in optical tweezer\cite{paper12}, microscopy\cite{paper06}, optical communication\cite{paper13}, astronomy\cite{paper14} and crystallography\cite{paper15}, etc. 
	
    The generation of vortices is the key to the research of science and technology involving vortices. There are various methods for generating vortex beams such as spiral phase plates\cite{paper16}, computer-generated holograms\cite{paper17}, metasurfaces\cite{paper18}, mode converters\cite{paper19,paper20}. It is worth noting that the aforementioned methods always require specific optical components and techniques, however, the water can easily produce vortices as it flows over a simple circular aperture. Since there are many similar definitions of optical and water vortices\cite{paper23,paper24,paper25,paper26}, such as OAM density and vorticity, OAM and eddy flux, it is worth investigating whether a light beam can also easily produce vortex when it is diffracted by a circular aperture.
	
    Here, we study the aperture diffraction of light and show that the light field in the Fresnel region of the diffracted circularly polarized beam carries OAM. Our research can be applied to develop a new way for generating optical vortices, which can be used to trap nanoparticles and make orbital rotation. Furthermore, circular aperture diffraction is a basic diffraction phenomenon of light. So our study can also lead to a deeper understanding of the optical diffraction. 

    Consider a left-hand circularly polarized Gaussian beam diffracted by a circular aperture with radius $a$ located at the plane $z = 0$ in the Cartesian coordinate system, where the $z$ axis is taken to be the propagation axis. According to the vectorial Rayleigh-Sommerfeld diffraction integral\cite{paper32,paper33}, the diffracted light can be expressed as
	\begin{equation}
	\begin{aligned}
	\begin{split}
	E_x(\rm \boldsymbol{\rho}) =&  -\emph i\,\frac{\emph k\emph z{\rm exp}(\emph {ik} \rho)}{\rho^2}
	\rm E_0\sum_{\emph m=0}^{+\infty} \frac{\emph k^{2\emph m}}{2^{2\emph m+1}(\emph m!)^2}\\
	&\times\frac{\emph h^{2\emph m}}{\emph g^{\emph m+1}}[\Gamma(1+\emph m,-\emph a^2\emph g)-\emph m!]
	\end{split}
	\end{aligned}
	\end{equation}
	\begin{equation}
	\begin{aligned}
	\begin{split}
	E_y(\rm \boldsymbol {\rho}) =&  \frac{\emph k\emph z{\rm exp}(\emph {ik} \rho)}{\rho^2}
	\rm E_0\sum_{\emph m=0}^{+\infty} \frac{\emph k^{2\emph m}}{2^{2\emph m+1}(\emph m!)^2}\\
	&\times
	\frac{\emph h^{2\emph m}}{\emph g^{\emph m+1}}[\Gamma(1+\emph m,-\emph a^2\emph g)-\emph m!]
	\end{split}
	\end{aligned}
	\end{equation}
	\begin{equation}
	\begin{aligned}
	\begin{split}
	E_z(\rm \boldsymbol {\rho}) =&{\rm exp}({i \theta} )
	\frac{k{\rm exp}(\emph {ik} \rho)}{\rho^2}
	\rm E_0
	\{
	\emph i\sum_{\emph m=0}^{+\infty} \frac{\emph k^{2\emph m}}{2^{2\emph m+1}(\emph m!)^2}\\
	&\times\frac{\emph h^{2\emph m}}{\emph g^{\emph m+1}}[\Gamma(1+\emph m,-\emph a^2\emph g)-\emph m!]\\
	&+\frac{1}{\rho}
	\sum_{\emph m=0}^{+\infty} \frac{\emph k^{2\emph m+1}}{2^{2\emph m+2}(\emph m!)(\emph m+1)!}\\
	&\times
	\frac{\emph h^{2\emph m}}{\emph g^{\emph m+2}}[\Gamma(2+\emph m,-\emph a^2\emph g)-(\emph m+1)!] 
	\}
	\end{split}
	\end{aligned}
	\end{equation}
    where $k$ is the wave number, ${\rm \boldsymbol {\rho}} = x{\rm \textbf e}_x + y{\rm \textbf e}_y +  z{\rm \textbf e}_z$, \textbf e$_x$, \textbf e$_y$ and \textbf e$_z$ are the unit vectors along the $x$, $y$ and $z$ directions, respectively, $\Gamma(\cdot)$ is the incomplete Gamma function, $g=ik/(2\rho)-1/(4\omega_0^2)$, $h=r/\rho$ and $r={\sqrt{x^2+y^2}}$. Note that equations (1) to (3) are valid in Fresnel and Fraunhofer diffraction regions. According to equation (3), there exists a helical phase term exp$(i\theta)$ in the $z$ component, indicating that the diffracted light carries OAM and the topological charge is $l$ = 1. 
	\begin{figure}[h!]
		\centering\includegraphics[width=1\linewidth]{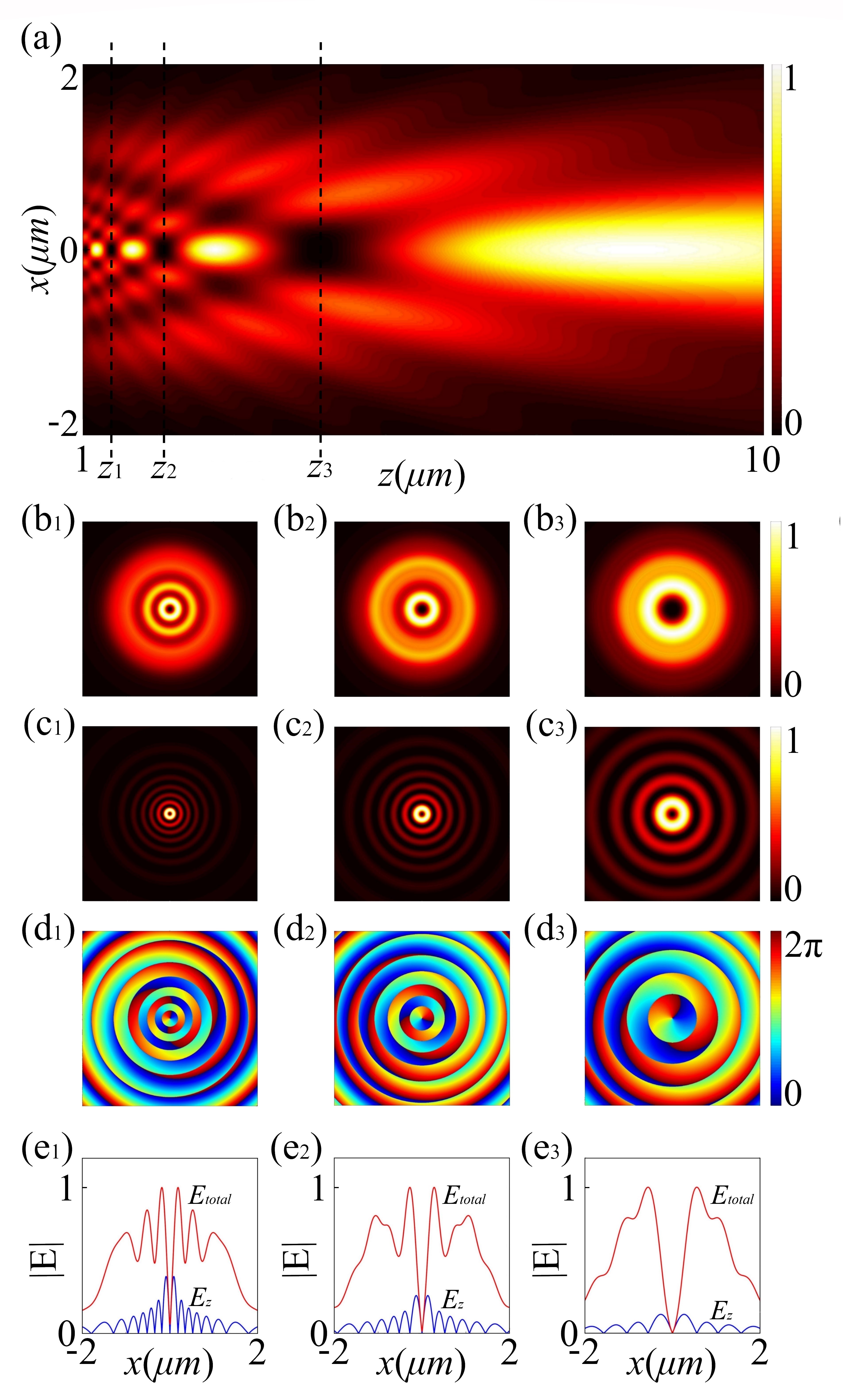}
		\caption{Theoretical intensity and phase distributions of the light field diffracted by a circular aperture according to equations (1)-(3). (a) The intensity distribution of the diffracted light in $xoz$ plane. (b) The total intensity distributions in planes $z=z_1$ (b$_1$), $z_2$ (b$_2$) and $z_3$ (b$_3$). (c) The intensity distributions of longitudinal component in planes $z=z_1$ (c$_1$), $z_2$ (c$_2$) and $z_3$ (c$_3$). (d) The phase distributions of longitudinal component in planes $z=z_1$ (d$_1$), $z_2$ (d$_2$) and $z_3$ (d$_3$). (e)  Cross sections of the total electric field and the longitudinal component along the $x$ direction in planes $z=z_1$ (e$_1$), $z_2$ (e$_2$) and $z_3$ (e$_3$).}
	\end{figure}
	
    Based on equations (1) to (3), Fig. 1 shows simulations of the light field behind a circular aperture under illumination of a left-hand circularly polarized Gaussian beam, where the intensity distributions are given by $I = |E_x|^2 + |E_y|^2 + |E_z|^2$. The wavelength of incident light in vacuum is  $\lambda$ = 532 $n$m, the waist width is $\omega_0$ = 0.7 $m$m and the radius of the circular aperture is $a$ = 1.8 $\mu $m. Figs. 1(b)-(d) show the intensity and phase distributions of the diffraction light field in planes $z = z_1, z_2$ and $z_3$, respectively. It can be obtained that, the intensity distributions of longitudinal component $I_z$ are hollow, which is caused by the helical phase distributions with topological charge $l$ = 1 [Fig. 1(d)]. According to Fig. 1(e), the longitudinal component of the diffracted light is obvious and non-negligible in circular aperture diffraction, meaning that the light field carries a certain amount of OAM. According to the Noether theorem, the total longitudinal AM J$_z$ carried by the light field is conserved. However, the longitudinal SAM S$_z$ decreases due to the circular aperture diffraction\cite{paper34}. It indicates that part of the longitudinal SAM S$_z$ is transferred to the longitudinal OAM L$_z$ of the diffracted light, which is caused by the spin-orbit interaction\cite{paper27}. 
	
	Next, the finite difference time domain (FDTD) method is used to verify the above theoretical results, in which the material of the circular aperture screen is perfect electric conductor. The simulation results based on FDTD method are shown in Fig. 2. Consisting with the above theoretical results, we can get a conclusion that the light field carries a certain amount of OAM converted from SAM. 
	
	\begin{figure}[h!]
		\centering\includegraphics[width=1\linewidth]{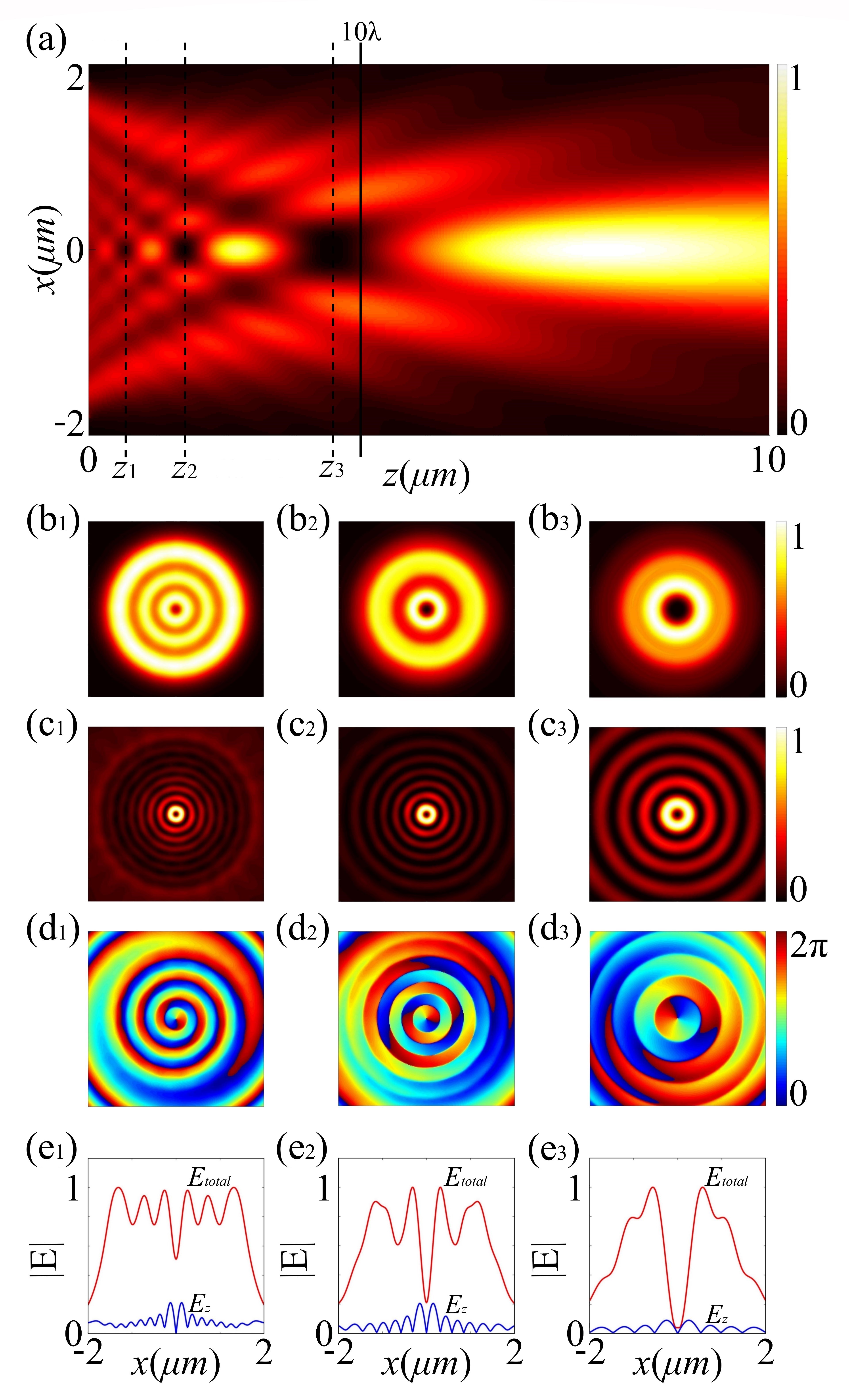}
		\caption{Simulation of the intensity and phase distributions of the light field diffracted by a circular aperture with FDTD method. (a) The intensity distribution of the diffracted light in plane $y$ = 0. (b) The total intensity distributions in planes $z=z_1$ (b$_1$), $z_2$ (b$_2$) and $z_3$ (b$_3$). (c) The intensity distributions of longitudinal component in planes $z=z_1$ (c$_1$), $z_2$ (c$_2$) and $z_3$ (c$_3$). (d) The phase distributions of longitudinal component in planes $z=z_1$ (d$_1$), $z_2$ (d$_2$) and $z_3$ (d$_3$). (e)  Cross sections of the total electric field and the longitudinal component along the $x$ direction in planes $z=z_1$ (e$_1$), $z_2$ (e$_2$) and $z_3$ (e$_3$).}
	\end{figure}
	
	It is a common phenomenon that small objects can be trapped in water vortices. Therefore, we study the trapping effect of the diffracted light on metal nanoparticles and demonstrate that the diffracted light carries certain OAM. Let us consider a spherical gold nanoparticle in water with a radius $d$ much smaller than the wavelength of the incident light. The optical force on the particle can be written as\cite{paper36}
	\begin{equation}
	\begin{split}
	\left \langle \rm \textbf F \right \rangle = &\frac{1}{4}\varepsilon\varepsilon_0\rm Re\left\{\alpha\right\}\nabla{\left|\rm \textbf E\right|}^2+\frac{\sigma \emph n}{2\emph c}\rm Re\left\{\textbf E\times\textbf H^\ast\right\}\\
	&+\frac{\sigma \emph c}{\emph n}\nabla\times\left\{\frac{\varepsilon\varepsilon_0}{4\omega \emph i}\textbf E\times\textbf E^\ast\right\}
	\end{split}
	\end{equation}
	where $c$ is the speed of light in vacuum, $\omega$ is the angular frequency of the light field, $\varepsilon$ and $n$ are the relative permittivity and refraction index of water, respectively. $\alpha$ is the complex polarizability of the spherical gold nanoparticle. $\sigma=k\rm Im\left\{\alpha\right\}$ denotes the extinction cross section of the particle. The first term denotes the gradient force \textbf F$_1$. The second term can be easily identified as the radiation force \textbf F$_2$. The third term is the spin curl force  \textbf F$_3$. For particles with radii much smaller than the wavelength of the incident light, the complex polarizability can be written as
	\begin{equation}
	\alpha = \frac{\alpha_0}{1-i\alpha_0k^3/(6\pi)},\quad \alpha_0 = 4\pi d^3\frac{\varepsilon_{\rm p}(\omega)-\varepsilon}{\varepsilon_{\rm p}(\omega)+2\varepsilon}
	\end{equation}
	where $\varepsilon_{\rm p}(\omega)$ is the relative permittivity of the spherical gold nanoparticle. In the case of $\lambda=532$ $n$m, the relative permittivity $\varepsilon_{\rm p}=-5.45+2.2i$. 
	
	\begin{figure}[h!]
		\centering\includegraphics[width=1\linewidth]{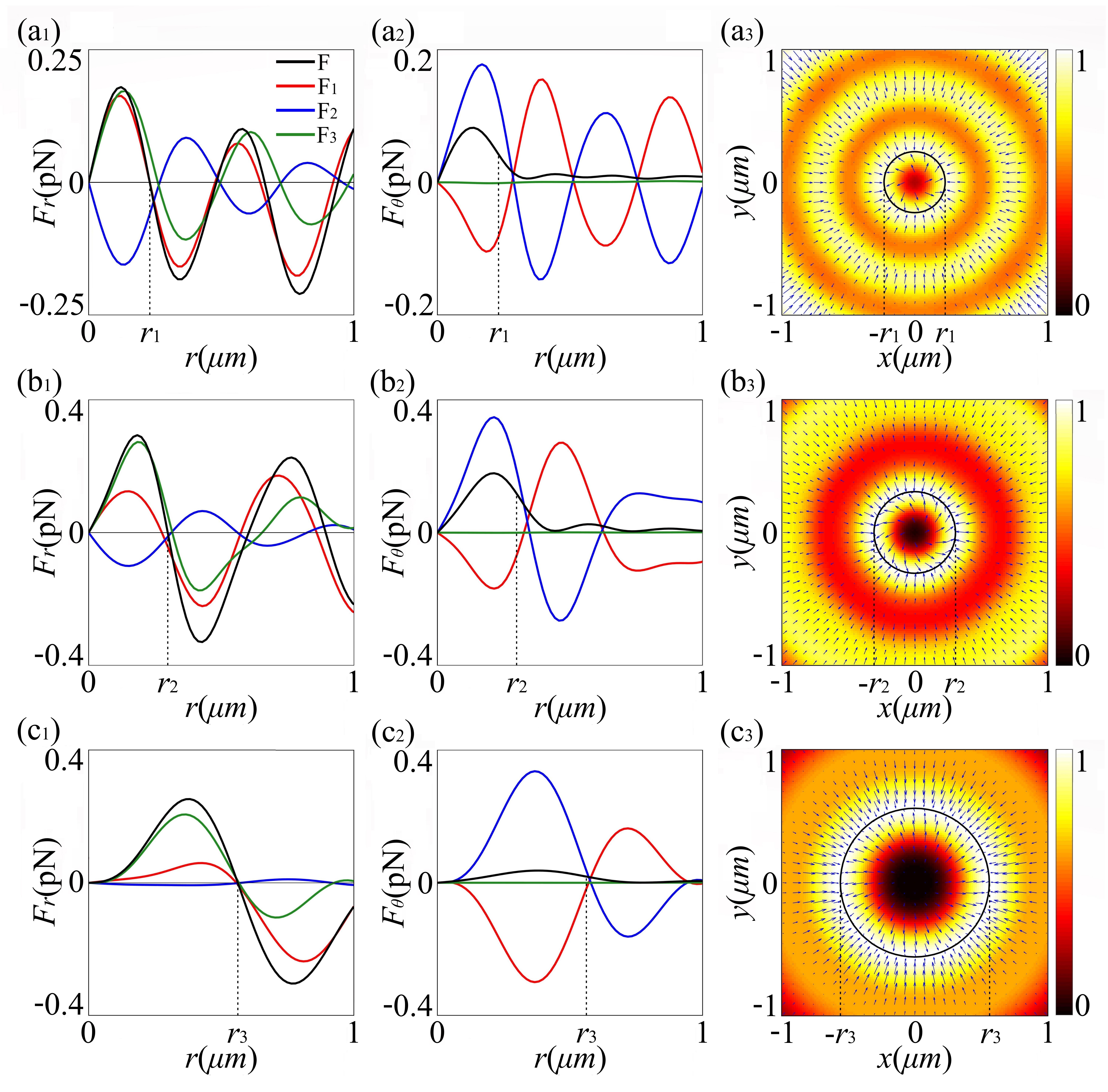}
		\caption{Simulation of the optical forces on the spherical gold particle with radius $d=50nm$ in the diffracted light field. (a), (b), (c) the optical forces on the particles in the planes $z =z_1$ (a), $z_2$ (b) and $z_3$ (c). ($\rm a_1$), ($\rm b_1$), ($\rm c_1$) Cross sections of radial components of optical forces. ($\rm a_2$), ($\rm b_2$), ($\rm c_2$) Cross section of angular components of optical forces, where the positive direction is counterclockwise. ($\rm a_3$), ($\rm b_3$), ($\rm c_3$) the vector distributions of the total optical forces, where the background is the intensity distributions of the diffracted light and the black circles are the orbits of the particles rotating around the beam axis.}
	\end{figure}
	
	The optical forces on the particles are shown in Fig. 3. Here, the incident optical power density is $\rm P\sim5$ $\rm mW/\mu m^2$\cite{paper36}. Note that this power density is equivalent to that obtained by focusing a 100mW beam into a spot with a diameter of 5$\mu m$. Based on the equation $(n/c)\rm P=\varepsilon_0\varepsilon {E_0}^2/2$, the electric field amplitude $\rm E_0$ can be calculated. The radial components of the optical forces on the particles are presented in Figs. 3($\rm a_1$), ($\rm b_1$), ($\rm c_1$) and the angular components are presented in Figs. 3($\rm a_2$), ($\rm b_2$), ($\rm c_2$). From Figs. 3($\rm a_1$), ($\rm b_1$) and ($\rm c_1$), under the combined action of the gradient forces \textbf F$_1$, radiation forces \textbf F$_2$ and spin curl force  \textbf F$_3$, the particles in the planes $z =z_1$, $z_2$ and $z_3$ will be trapped at the radial positions \textit{r} $=r_1$, $r_2$ and $r_3$, respectively. Meanwhile, from Figs. 3($\rm a_2$), ($\rm b_2$) and ($\rm c_2$), the radially trapped particles are subjected to the counterclockwise angular forces. The forces that rotate the particles around the beam axis are derived from the radiation forces \textbf F$_2$ and spin curl force \textbf F$_3$. Furthermore, particles at any position in the planes $z =z_1$, $z_2$ and $z_3$ are subjected to the non-zero counterclockwise angular forces. The vector distributions of the total optical forces is exhibited in Fig. 3($\rm a_3$), ($\rm b_3$) and ($\rm c_3$). 
	
	We know that the particles are driven not only by the optical forces but also by the random collisions of water molecules called Brownnian motion. To trap and rotate the particles, the optical forces must be able to overcome Brownian motion. The motion of a spherical gold nanoparticle with radius $d$ (negligible mass) driven by optical forces and the collisions from water molecules can be described by Langevin equation\cite{paper37,paper38,paper39}
	\begin{equation}
	\gamma \frac{d\textbf r}{dt}= \textbf F(\textbf r)+\sqrt {2\gamma k_{\rm B}\rm T}\cdot W(t)
	\end{equation}
	where $\gamma=6\pi\eta d$ is the friction coefficent, $\eta$ is the viscosity of medium, $k_{\rm B}$ is Boltzmann constant. For water at absolute temperature $\rm T=298$ K, $\eta=0.89\times10^{-3}$ $k\rm g/(\rm m\cdot \rm s)$. $W(t)$ is the white noise term used to model random collisions from water molecules. 
	
	\begin{figure}[h!]
		\centering\includegraphics[width=1\linewidth]{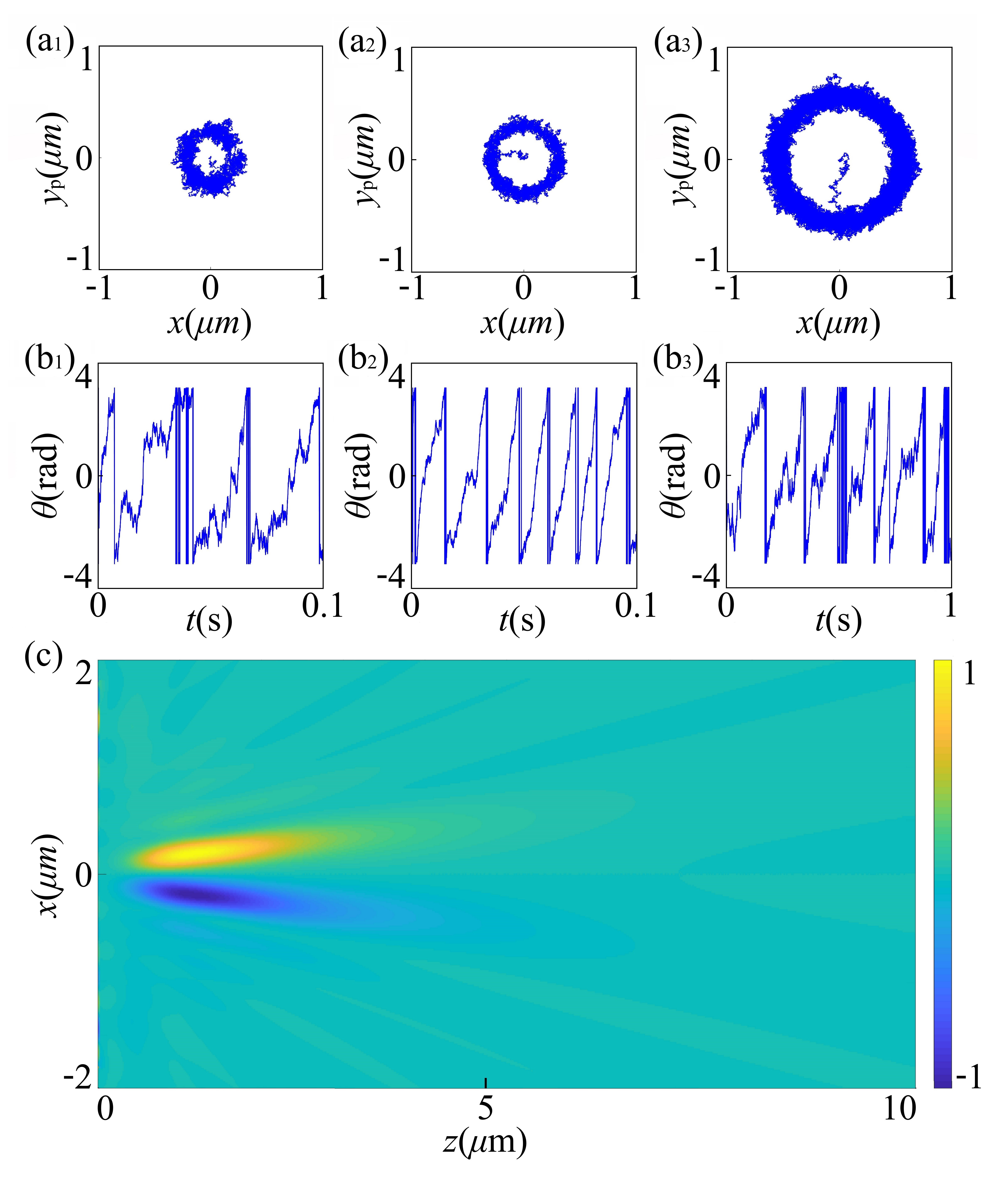}
		\caption{(a) The motion trajectories of particles with radius $d=50$ $n$m starting from position $(0,0,z)$ in planes $z =z_1$ (a$_1$), $z_2$ (a$_2$) and $z_3$ (a$_3$). (b) The azimuthal positions of particles in the planes $z =z_1$ (b$_1$), $z_2$ (b$_2$) and $z_3$ (b$_3$) vary with time $t$. (c) The azimuthal optical forces on particles in $xoz$ plane, where the positive direction is $+y$ direction.}
	\end{figure}
	
	We simulate the motion trajectories of particles with radius $d=50$ $n$m starting from position $(0,0,z)$ in different $z$ planes, where the time step is set as 50 $n$s. According to Fig. 4$(\rm a)$, the trajectories of particles in planes $z =z_1$, $z_2$ and $z_3$ are rings, which is consistent with the simulation results in Fig. 3. The evolution of the circular positions of the particles are presented in Fig. 4$(\rm b)$. It can be obtained that the particles rotate around the beam axis counterclockwise. Moreover, since the incident light is left-circularly polarized, it carries a SAM of $\hbar$ per photon. Due to diffraction of light, the SAM carried by the light field is reduced. According to Noether theorem, the OAM carried by the light field increases. So the diffracted light carries positive OAM. This means that the particles in the diffracted light will rotate counterclockwise around the beam axis. The analysis is in accord with the simulation results in Fig. 4(b). In fact, not only the diffracted light in planes $z =z_1$, $z_2$ and $z_3$ carries OAM, but also the diffracted light in any $z$ plane carries a certain amount of OAM. The reason is that there will be no spin-orbit interaction in the light field without the diffraction of the beam itself and the external action and there will be no conversion of SAM and OAM. Therefore, the OAM carried by the diffracted light will also remain constant. That is, the particles located at almost any position in the diffracted light always subject to a counterclockwise azimuthal force as expressed in Fig. 4(c). 
	
	In conclusion, we have demonstrated the spin-to-orbital AM conversion in the circular aperture diffraction of a circularly polarized beam. The reduced longitudinal SAM is converted into OAM carried by the diffracted light because of the spin-orbit interaction. Then we analyzed the optical forces on the spherical gold nanoparticles in the diffracted light field, which further proved that the diffracted light carries a certain amount of OAM. Nanoparticles can be rotated around the beam axis by the angular scattering force from the spin-orbit coupling and the spin curl force. Our study can be applied to develop new methods for generating vortex beam and explore new nanoscale optical manipulation techniques, and we believe it will also contribute to a deeper understanding of the diffraction.\\
	
	This work was supported by the National Natural Science Foundation of China (Nos. 11874102 and 12174047), Sichuan Province Science and Technology Support Program (No. 2020JDRC0006).
	
	\bibliographystyle{apsrev4-2}

\end{document}